\documentclass[%
 reprint,
superscriptaddress,
amsmath,amssymb,
aps,
prb
]{revtex4-2}

\usepackage{graphicx}
\usepackage{dcolumn}
\usepackage[latin1]{inputenc}
\usepackage{bm}
\usepackage[colorlinks,linkcolor=blue,anchorcolor=blue,citecolor=blue,urlcolor=blue]{hyperref}
\usepackage{xcolor}
\usepackage[mathlines]{lineno}

\DeclareMathOperator{\sign}{sign}
\renewcommand{\Im}{\operatorname{Im}}

\begin{document}

\title{Intrinsic thermal Hall effect of optical phonons enhanced by discrete rotational symmetry}

\author{Xuesong Hu}

\affiliation{International Center for Quantum Materials, Peking University, Beijing 100871, China}
\author{Junren Shi} \email{junrenshi@pku.edu.cn}

\affiliation{International Center for Quantum Materials, Peking University, Beijing 100871, China}

\affiliation{Collaborative Innovation Center of Quantum Matter, Beijing 100871, China}

\date{\today}

\begin{abstract}
    We investigate the intrinsic thermal Hall conductivity contributed by optical phonons in a cubic system.  The discrete rotational symmetry of the system splits the degeneracy of transverse modes across most regions of wave-vector space,  except along a few high-symmetry lines.  Consequently, in the presence of an external magnetic field, phonon Berry curvatures become sharply peaked near these high-symmetry lines.  We find that the singular distribution of the Berry curvature induces an intrinsic thermal Hall conductivity that is significantly enhanced compared to an isotropic system.  It exhibits a nonlinear $B\ln B$ dependence on the magnetic field $B$ and a non-monotonic temperature dependence.  At elevated temperatures, it reverses sign and approaches a non-vanishing value asymptotically.  Our analysis indicates that the behavior results from competition between contributions from Berry curvatures near different high-symmetry lines.
\end{abstract}

\maketitle


\section{\label{sec:level1}Introduction}

In recent years, thermal Hall conductivity measurements have attracted significant interest because of their ability to elucidate the properties of neutral excitations~\cite{strohm_phenomenological_2005,inyushkin_phonon_2007,onose_observation_2010,ideue_effect_2012,hirschberger_large_2015,sugii_thermal_2017,ideue_giant_2017,liu_magnon_2017,kasahara_unusual_2018,lefrancois_evidence_2022}. Analogous to the electric Hall effect, the thermal Hall effect occurs when a longitudinal heat current induces a transverse temperature gradient in the presence of a perpendicular magnetic field~\cite{strohm_phenomenological_2005,inyushkin_phonon_2007}. Large thermal Hall conductivities have been recently observed in various magnetic~\cite{grissonnanche_giant_2019,grissonnanche_chiral_2020,boulanger_thermal_2020,zhang_anomalous_2021,lefrancois_evidence_2022,chen_large_2022,xu_thermal_2023,ataei_phonon_2024} and non-magnetic insulators~\cite{li_phonon_2020,li_phonon_2023}, where the heat carriers are neutral quasiparticles such as magnons or phonons. This phenomenon is especially intriguing for non-magnetic insulators, as the heat carriers in these materials are likely phonons, which are charge neutral and do not couple directly to the magnetic field~\cite{sheng_theory_2006,kagan_anomalous_2008,mori_origin_2014}.

Theoretical approaches to understanding the effect can be categorized into intrinsic and extrinsic mechanisms.  The intrinsic mechanism focuses on the topological properties of heat-carrying quasi-excitations, such as spinons~\cite{samajdar_thermal_2019,zhang_thermal_2024}, magnons~\cite{mook_magnon_2014,murakami_thermal_2017,zhang_thermal_2024}, and phonons~\cite{zhang_topological_2010,Taoqin2012,chen_enhanced_2020,ye_2021_phononhall,zhang_phonon_2021,mangeolle_phonon_2022}.  The mechanism may explain satisfactorily the effect in magnetic insulators where excitations (e.g., spinons and magnons) can either couple directly to the magnetic field~\cite{murakami_thermal_2017,zhang_anomalous_2021,xu_thermal_2023,samajdar_thermal_2019,mook_magnon_2014,zhang_thermal_2024} or indirectly via spin-phonon coupling~\cite{bonini_frequency_2023,xu_2024}.  However, it fails to explain the large thermal Hall conductivities observed in non-magnetic insulators, as the phonon-magnetic field coupling, which originates from Lorentz forces on charged ions, is characterized by a tiny energy scale set by ion cyclotron energies $\sim 10^{-5}\, \mathrm{meV}/\hbar$ at $B=10\mathrm{T}$.  On the other hand, the extrinsic mechanism attributes thermal Hall effects to quasiparticle scattering by imperfections such as structural domains, impurities and defects.  In particular, it is argued that skew scattering may account for the large thermal Hall conductivities as its contributions increase with the mean free path of phonons~\cite{mori_origin_2014,chen_enhanced_2020,guo_extrinsic_2021, flebus_charged_2022}.

Prior studies estimate the intrinsic contribution to the thermal Hall conductivity using a formula for isotropic acoustic phonon modes in the long-wavelength limit~\cite{Taoqin2012,chen_enhanced_2020}. The approach may underestimate the intrinsic contribution, especially at high temperatures when optical modes can be excited.  Unlike acoustic phonons, which have vanishing coupling strength to a magnetic field in the long-wavelength limit due to charge neutrality and translational symmetry, optical phonons can couple  more strongly to a magnetic field as they are not bound by these constraints.  Moreover, crystal structures of real materials reduce the continuous rotational symmetry to discrete ones, which may fundamentally alter phonon bands and their topological properties.  Therefore,  estimating the intrinsic contribution using an  isotropic acoustic phonon model is inadequate.  A theoretical investigation into the contribution of optical phonons, considering properly the effect of discrete rotational symmetry in real materials, is necessary.

In this paper, we analyze the thermal Hall conductivity contributed by optical phonons using a continuous effective model for a lattice of polar molecules.  We incorporate the discrete rotational symmetry of a cubic system in the effective model to account for the effect of the crystal lattice.  The discrete symmetry splits the degeneracy of transverse optical phonon modes, fundamentally altering the distribution of phonon Berry curvatures, from a continuous one to that concentrated near a few specific high-symmetry lines in wave-vector space.  The redistribution of the Berry curvatures results in an enhanced thermal Hall conductivity.  Notably, the contribution exhibits a nonlinear $B\ln B$ dependence on the external magnetic field $B$, as well as a non-monotonic temperature dependence.  At elevated temperatures, it reverses sign and approaches a non-vanishing value asymptotically.  Our analysis reveals that this behavior is a result of competing contributions from regions near different high-symmetry lines.

The remainder of the paper is organized as follows.  In Sec.~\ref{sec:Bc}, we introduce our model.  In Sec.~\ref{sec:bc}, we analyze the structure and Berry curvatures of optical phonon bands and the effect of the discrete rotational symmetry.  Based on the analysis, we determine the intrinsic thermal Hall conductivity contributed by optical phonons in Sec.~\ref{sec:therm-hall-cond}, where we also provide a theoretical understanding of its general behavior.  Finally, we summarize and discuss our findings in Sec.~\ref{sec:summary}.  Details of derivations are provided in Appendices.

\section{Model}\label{sec:Bc}

Our study employs a continuous elastic model to describe the long-wavelength vibrational modes of a cubic crystal lattice of polar molecules.  We model the polar molecules as electric dipoles.  They interact via local elastic forces and the long-range Coulomb interaction.

The Lagrangian density of the continuous elastic model is written as~\cite{landua_elasticity,chen_enhanced_2020}
\begin{multline}
\mathcal{L}=\frac{\rho}{2}\dot{\bm{u}}^{2}-\frac{\rho\omega_{0}^{2}}{2}\bm{u}^{2}-\frac{\rho\lambda_{1}}{2}(\boldsymbol{\nabla}\cdot\bm{u})^{2}-\frac{\rho\lambda_{2}}{2}(\boldsymbol{\nabla}\bm{u})^{2}\\
-\frac{\rho\lambda_{3}}{2}\sum_{i}(\partial_{i}u_{i})^{2}-\frac{\rho_{e}^{\prime}\bm{B}}{2}\cdot(\bm{u}\times\dot{\bm{u}})\\
-\frac{\rho_{e}^{2}}{4\pi\epsilon_{\infty}}\int d^{3}\boldsymbol{r}'\frac{\boldsymbol{\nabla}\cdot\bm{u}(\boldsymbol{r})\boldsymbol{\nabla}'\cdot\bm{u}(\boldsymbol{r}')}{|\boldsymbol{r}-\boldsymbol{r}'|},\label{eq:lagrangian}
\end{multline}
where $\bm{u}$ represents the displacement vector of the dipoles, $\rho$ denotes the reduced mass density, and $\omega_{0}$ is the local vibrational frequency, corresponding to the frequency of transverse optical phonon modes at the $\Gamma$ point.  The parameters $\lambda_{i}$, $i=1,2,3$, are elastic moduli for characterizing the local elastic energy~\cite{landua_elasticity}.  For a cubic system, in addition to the usual isotropic bulk and shear moduli from $\lambda_{1}$ and $\lambda_{2}$, an extra elastic modulus $\lambda_{3}$ breaks the continuous rotational symmetry.   The rest of the Lagrangian density describes the coupling of the electric dipoles to an external magnetic field $\bm B$ and the Coulomb interaction between them, with $\rho_{e}^{\prime}$ and $\rho_{e}$ representing charge densities that define the coupling and interaction strengths~~\cite{chen_enhanced_2020}.  For diatomic polar molecules, these parameters are determined from the masses and electric charges of ions by $\rho=m_{+}m_{-}/\Omega(m_{+}+m_{-})$, $\rho_{e}=q/\Omega$, and $\rho_{e}^{\prime}=q(m_{+}-m_{-})/\Omega(m_{+}+m_{-})$, where $m_{+}$ ($m_{-}$) is the mass of positive (negative) ions,  $q$ is the positive ion charge, and $\Omega$ is the volume of the unit cell of the system.

The continuous model can be generalized to systems with other discrete rotational symmetries by adjusting the elastic energy part of the Lagrangian.  Different symmetries will have different symmetry-broken terms and moduli, as discussed in Ref.~\onlinecite{landua_elasticity}.

\section{Phonon bands and Berry curvature}\label{sec:bc}

\subsection{Phonon bands}
 \label{sec:phonon-bands}
The band structure of phonons can be determined by solving the generalized eigen-equation, $\omega_{ki}\psi_{\boldsymbol{k}i}=\Tilde{H}_{\boldsymbol{k}}\psi_{\boldsymbol{k}i}$, in the wave-vector space with~\cite{Taoqin2012}
\begin{equation}
\tilde{H}_{\bm{k}}\equiv\begin{bmatrix}0 & \mathrm{i}I_{3\times3}\\
-\mathrm{i}D(\bm{k}) & \mathrm{i}G
\end{bmatrix},
\label{Hktilde}
\end{equation}
where $\psi_{\bm{k}i}$ denotes the six-component eigenvector of the equation, and can be decomposed as $\psi_{\bm{k}i}=\left[\bm{u}_{\bm{k}i}, -\mathrm{i}\omega_{\bm{k}i}\bm{u}_{\bm{k}i} \right]^{T}$ with $\bm{u}_{\bm{k}i}$ representing the displacement vector of the dipole field. $G$ and $D(\bm k)$ are $3\times3$ matrices with the matrix elements
\begin{align}
  G_{\alpha\beta} & =-\varepsilon_{\alpha\beta z}\omega_{B},\\
D_{\alpha\beta}(\bm{k}) & =\left(\omega_{0}^{2}+\lambda_{2}k^{2}+\lambda_{3}k_{\alpha}^{2}\right)\delta_{\alpha\beta}\nonumber \\
 & +\left(\lambda_{1}+\frac{\alpha^{2}}{k^{2}}\right)k_{\alpha}k_{\beta},
\end{align}
with $\alpha=\rho_{e}/\sqrt{\rho\epsilon_{\infty}}$, and $\omega_{B}=\rho_{e}'B/\rho$ being the ionic cyclotron frequency, which sets the energy scale of the coupling between optical phonons and the external magnetic field.  We assume that the magnetic field is along the $z$ direction.
    
\begin{figure}[t]
\includegraphics[width=\columnwidth]{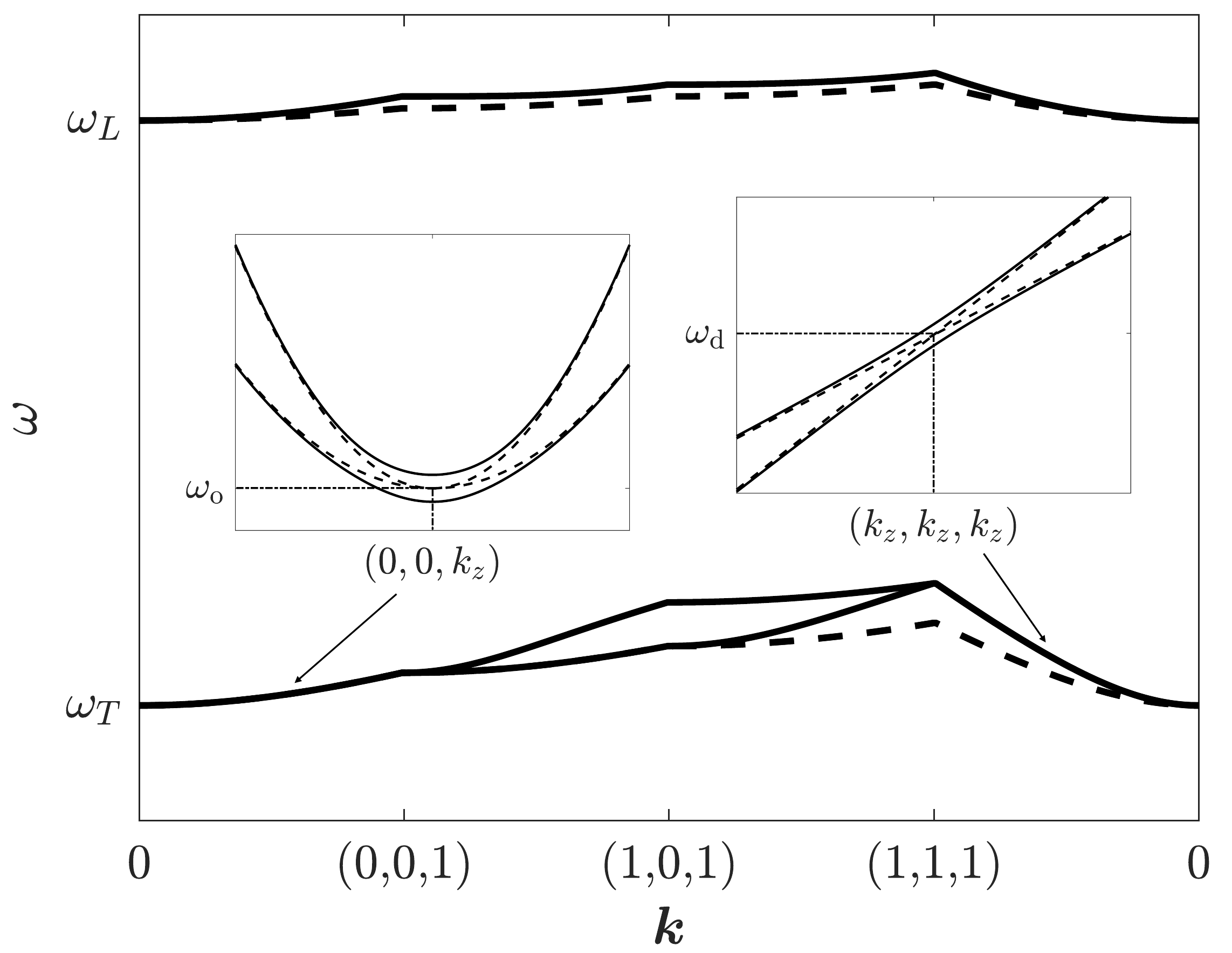}
  \caption{Phonon dispersions along selected directions in wave-vector space.  Solid (dashed) lines represent dispersions for $\lambda_{3}\ne 0$ ($\lambda_{3}=0$).  Insets show the splitting of degeneracy near points $(0,0,k_{z})$ and $(k_{z},k_{z},k_{z})$ in the constant $k_{z}$ plane when an external magnetic field is applied along the $z$ direction, with the solid (dashed) line representing the dispersions for $B\ne 0$ ($B=0$). }
  \label{fig:band}
\end{figure}

Figure~\ref{fig:band} shows phonon dispersions.  There are three phonon branches.  For an isotropic system ($\lambda_{3}=0$) in the absence of the magnetic field, they include two degenerate transverse modes as well as a longitudinal mode which is elevated to higher frequencies by the long-range Coulomb interaction.  At the $\Gamma$ point of the wave-vector space, we have the transverse phonon frequency $\omega_{T}=\omega_{0}$, and the longitudinal frequency $\omega_{L}=\sqrt{\omega_{0}^{2}+\alpha^{2}}$.  According to the LST relation~\cite{lyddane_polar_1941}, the ratio of $\omega_{L}$ to $\omega_{T}$ can be related to the  dielectric constant $\epsilon$: $\omega_{\text{L}}^{2}/\omega_{\text{T}}^{2}=\epsilon/\epsilon_{\infty}$.   In materials with large static dielectric constants, e.g., strontium titanate (STO)~\cite{yamada_neutron_1969,collignon_metallicity_2019}, $\omega_{L}$ is much larger than $\omega_{T}$. Therefore, in the subsequent analysis, we will ignore the contribution from the longitudinal mode to the thermal Hall coefficient and focus exclusively on the two low energy transverse modes.

The discrete rotational symmetry in a cubic system ($\lambda_{3}\ne 0$) breaks the degeneracy of the transverse modes.  As illustrated in Fig.~\ref{fig:band}, it splits the two transverse modes in most regions of the wave-vector space, except along the high-symmetry directions with $|k_{x}|=|k_{y}|=|k_{z}|$ as well as the $k_{x}$, $k_{y}$, $k_{z}$ axes. 

The degeneracy along these high-symmetry lines is further lifted when an external magnetic field is applied, as shown in the insets of Fig.~\ref{fig:band}. 

The splitting of the transverse modes near these high-symmetry  directions can be analyzed.  For the point $\bm{k}=(k_{z}+\delta k_{x},k_{z}+\delta k_{y},k_{z})$ near $\bm k_{0}\equiv (k_{z},k_{z}, k_{z})$, we expand $\Tilde{H}(\boldsymbol{k})$ to the linear order of $\delta k_{x}$ and $\delta k_{y}$,  and project it onto the subspace spanned by the two degenerate transverse mode eigenvectors at $\bm k_{0}$ which have the displacement vectors shown in Eq.~\eqref{eq:ud}.  The corresponding $2\times2$ effective Hamiltonian takes the form of the Dirac model (see Appendix~\ref{app:H_eff}):
\begin{equation}
\Tilde{H}_{\boldsymbol{k}}^{\mathrm{(d)}}\approx(\tau_{1}\delta k_{2}-\tau_{2}\delta k_{1})v_{\mathrm{d}}(k_{z})+\tau_{3}h_{\mathrm{d}}+\omega_{\mathrm{d}}(\boldsymbol{k}),\label{eq:H_eff_R}
\end{equation}
where $\delta k_{1}\equiv(\delta k_{x}-\delta k_{y})/\sqrt{2}$, $\delta k_{2}\equiv(\delta k_{x}+\delta k_{y})/\sqrt{6}$,   $\omega_{\mathrm{d}}(\bm{k})=\sqrt{\omega_{0}^{2}+(\lambda_{2}+\lambda_{3}/3)(k_{x}+k_{y}+k_{z})^{2}/3}$, and $\tau_{i}$ ($i=1,2,3$) are the Pauli matrices in the subspace. The coefficients of the Dirac model are determined by $h_{\mathrm{d}}\equiv-\omega_{B}/2\sqrt{3}$ and $v_{\mathrm{d}}(k_{z})\equiv\lambda_{3}k_{z}/\sqrt{6}\omega_{\mathrm{d}}(\bm{k}_{0})$.  The dispersion near $\bm k_{0}$ is thus given by
\begin{equation}                                                                \omega_{\bm{k},\pm}=\omega_{\mathrm{d}}(\boldsymbol{k})\pm\sqrt{(\delta k_{1}^{2}+\delta k_{2}^{2})v_{\mathrm{d}}(k_{z})^{2}+h_{\mathrm{d}}^{2}}.
  \label{eq:dispd}
\end{equation}
Effective Hamiltonians for other diagonal directions can be obtained by applying $C_{4}$ rotations around the $k_{z}$-axis.  Note that the superscript or subscript ``$\mathrm{d}$" in various quantities indicates their association with the regions near the diagonal direction, where $|k_{x}|=|k_{y}|=|k_{z}|$. It distinguishes them from similar quantities for the region near the $k_{z}$ axis.

The splitting of the degeneracy along the $k_{z}$ axis can be analyzed similarly.  At the point $\bm{k}=(\delta k_{x},\delta k_{y},k_{z})$ near $\bm k_{0}^{\prime}\equiv(0,0,k_{z})$, we expand $\Tilde{H}_{\boldsymbol{k}}$ to the quadratic order in $\delta k_{x}$ and $\delta k_{y}$.  Projecting the Hamiltonian onto the subspace  spanned by the two degenerate transverse mode eigenvectors at $\bm k_{0}^{\prime}$, which have displacement vectors given by Eq.~\eqref{eq:uo}, we can obtain the $2\times2$ effective phonon Hamiltonian (see Appendix~\ref{app:H_eff}):
\begin{equation}
\Tilde{H}_{\boldsymbol{k}}^{(\mathrm{o})}\approx\boldsymbol{\tau}\cdot\boldsymbol{d}(\boldsymbol{k})v_{\mathrm{o}}(k_{z})+\tau_{3}h_{\mathrm{o}}+\omega_{\mathrm{o}}(\boldsymbol{k}),\label{eq:H_eff_G}
\end{equation}
with $\omega_{\mathrm{o}}(\bm{k})\equiv\sqrt{\omega_{0}^{2}+\lambda_{2}k^{2}+\lambda_{3}(k_{x}^{2}+k_{y}^{2})}$, $\boldsymbol{d}(\boldsymbol{k})\equiv(\delta k_{x}^{2}-\delta k_{y}^{2},-\delta k_{x}\delta k_{y},0)$,  $h_{\mathrm{o}}\equiv-\omega_{B}/2$, and $v_{\mathrm{o}}(k_{z})\equiv\lambda_{3}/2\omega_{\mathrm{o}}(\bm k_{0}^{\prime})$.  The dispersions near $\bm k_{0}^{\prime}$ are
\begin{equation}
  \omega_{\bm{k},\pm}=\omega_{\mathrm{o}}(\boldsymbol{k})\pm\sqrt{d(\boldsymbol{k})^{2}v_{\mathrm{o}}(k_{z})^{2}+h_{\mathrm{o}}^{2}}.
  \label{eq:dispz}
\end{equation}
Here we use the superscript or subscript ``$\mathrm{o}$" in various quantities to indicates their association with the region near the origin of the constant $k_{z}$-plane.

\subsection{Berry curvature \label{sec:berry-curvature-}}

The Berry curvatures of the phonon bands can be determined using the definition $\boldsymbol{\Omega}_{\boldsymbol{k},i}=-\Im [\partial_{\boldsymbol{k}}\bar{\psi}_{\boldsymbol{k}i}\times \partial_{\boldsymbol{k}}\psi_{\boldsymbol{k}i}]$, where $\psi_{\boldsymbol{k}i}$ is the eigenvector obtained from the generalized eigen-equation and normalized by $\bar{\psi}_{\boldsymbol{k}i}\psi_{\boldsymbol{k}j}=\delta_{ij}$,  with $\bar{\psi}_{\boldsymbol{k}i}\equiv\psi_{\boldsymbol{k}i}^{\dagger}\Tilde{D}_{\boldsymbol{k}}$ and $\Tilde{D}_{\boldsymbol{k}}\equiv\left[\begin{smallmatrix} D_{\boldsymbol{k}} & 0\\ 0 & I_{3\times3} \end{smallmatrix}\right]$~\cite{Taoqin2012}.

\begin{figure}[t]
  \includegraphics[width=\columnwidth]{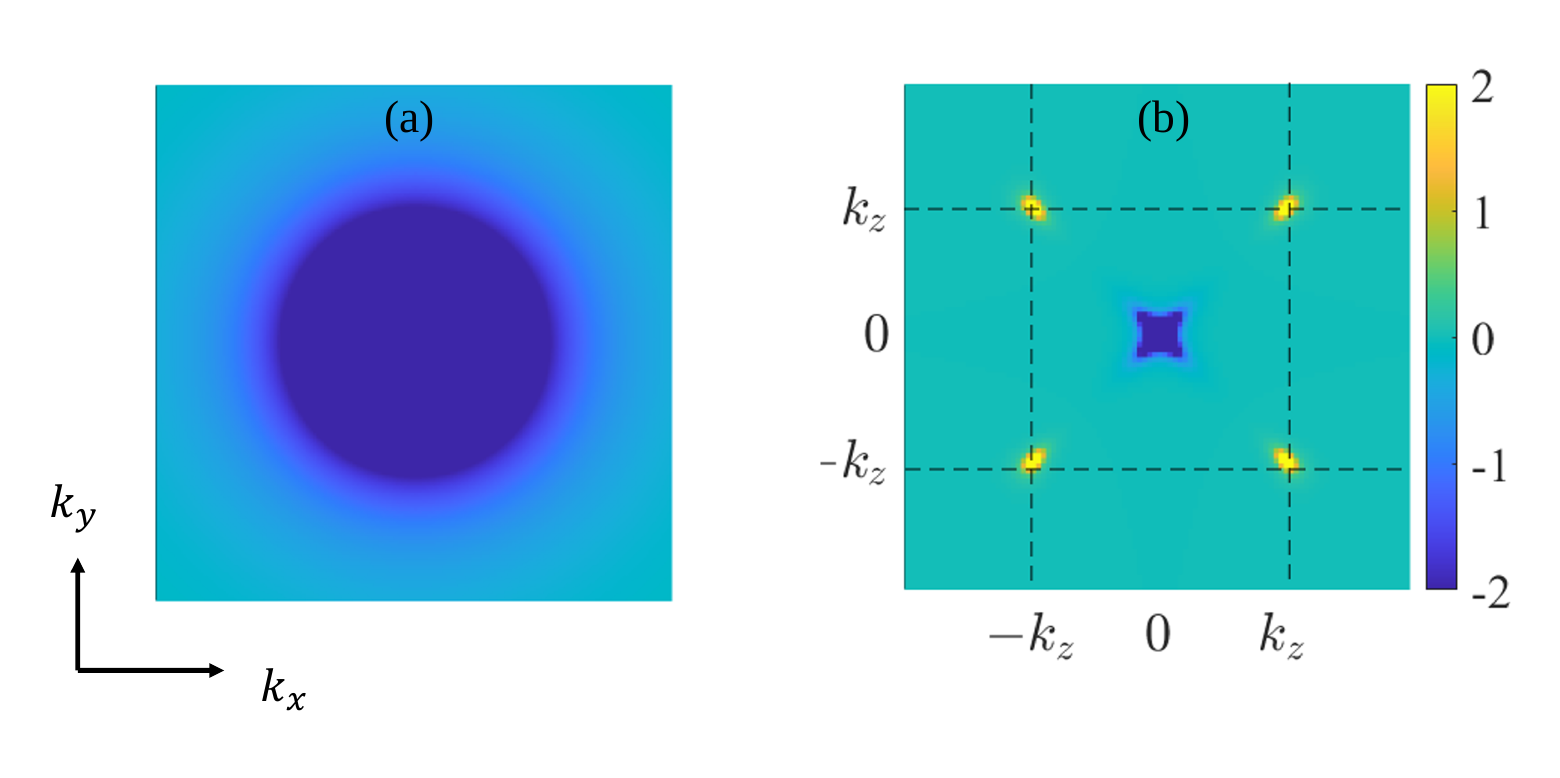}
  \caption{Berry curvature distribution of the upper transverse phonon band across a constant $k_{z}$ plane for (a) $\lambda_{3}=0$ and (b) $\lambda_{3}\ne 0$.  The parameters are $\lambda_{2}=1.5\times10^{4} \text{meV}^{2}\AA^{2}/\hbar^{2}$, $k_{z}=0.2 \AA^{-1}$ and $\omega_{B}=5 \times 10^{-5}$ meV$/\hbar$. For (b), we set $\lambda_{3}=\lambda_{2}$.}
  \label{fig:bc}
\end{figure}

We determine numerically the Berry curvatures of the two low-lying transverse modes in the presence of a magnetic field.  Figure~\ref{fig:bc} shows the Berry curvature distribution of the upper transverse phonon band across a wave-vector space cross-section at a constant $k_{z}$.  It is evident that the discrete rotational symmetry fundamentally alters the distribution.  In an isotropic system with $\lambda_{3}=0$, there is a single broad peak centered at the origin.  Conversely, when $\lambda_{3}\ne 0$, the center peak weakens, and additional sharp peaks emerge near the intersections of the diagonal high-symmetry lines with the $k_{z}$ plane.

The Berry curvatures near the high-symmetry lines can be determined approximately using the effective two-band models obtained in the last subsection.  For the region near $\bm{k}_{0}\equiv(k_{z},k_{z},k_{z})$, we employ the effective Hamiltonian Eq.~\eqref{eq:H_eff_R} and apply the general Berry-curvature formula for a two-band system~\cite{m_v_quantal_1984, qi_topological_2011}.  The Berry curvatures near $\bm{k}_{0}$, in a constant $k_{z}$ plane, are given by
\begin{equation}
\Omega_{\boldsymbol{k},\pm}^{z}\approx\pm\frac{1}{\sqrt{3}}\frac{q_{\mathrm{d}}}{\left[2(\delta k_{1}^{2}+\delta k_{2}^{2})+q_{\mathrm{d}}^{2}\right]^{3/2}},\label{eq:bc_R}
\end{equation}
for the upper ($+$) and lower ($-$) transverse phonon bands, respectively, with $q_{\mathrm{d}}\equiv\omega_{B}/\sqrt{6}|v_{\mathrm{d}}(k_{z})|$.  The $1/\sqrt{3}$ prefactor arises from the cosine of the angle between the $(111)$ direction and the direction of the magnetic field.  The Berry curvature peaks at $\bm k_{0}$ with a width set by $q_{\mathrm{d}}\propto\omega_{B}/\lambda_{3}$.  Since the ionic cyclotron frequency $\omega_{B}$ is tiny, the peak is sharp in the wave-vector space.   It contributes a total Berry phase of $\pm\pi\sign{(\omega_{B})}$.

Near the point $(0,0,k_{z})$, we make use of the effective Hamiltonian Eq.~\eqref{eq:H_eff_G}, which yields
\begin{equation}
\Omega_{\boldsymbol{k},\pm}^{z}\approx\mp\frac{q_{\mathrm{o}}^{2}}{\left(\left|\bm{d}(\bm{k})\right|^{2}+q_{\mathrm{o}}^{4}\right)^{3/2}}\left(\delta k_{x}^{2}+\delta k_{y}^{2}\right),\label{eq:bc_gamma}
\end{equation}
with $q_{\mathrm{o}}\equiv\sqrt{\omega_{B}/2v_{\mathrm{o}}(k_{z})}$ near the origin of the constant $k_{z}$-plane.  The Berry curvature contributes a total Berry phase of $\mp2\pi\sign{(\omega_{B})}$.


We see that the discrete symmetry introduces a fundamental change to the distribution of the Berry curvature.  It transforms the continuous distribution of an isotropic system to one with sharp peaks along the high-symmetry lines in the wave-vector space.  Because the widths of the peaks ($q_{\mathrm{d}}$ and $q_{\mathrm{o}}$) are set by $\omega_{B}/\lambda_{3}$ and $\omega_{B}\sim 10^{-5} \mathrm{meV}/\hbar$ (at $B=10\mathrm{T}$) is negligibly small, the value of $\lambda_{3}$ required to induce the change is much smaller than its actual values in real materials.  This indicates that an isotropic continuous effective model may not be adequate in describing the topological properties of phonon bands even in the long-wavelength limit.  Explicit consideration of crystal structure, as the $\lambda_{3}$ term does for a cubic system, is required. 

\section{Thermal Hall Conductivity}\label{sec:therm-hall-cond}

\subsection{Numerical results}

\begin{figure}[tb]
  \centering
 \includegraphics[width=\columnwidth]{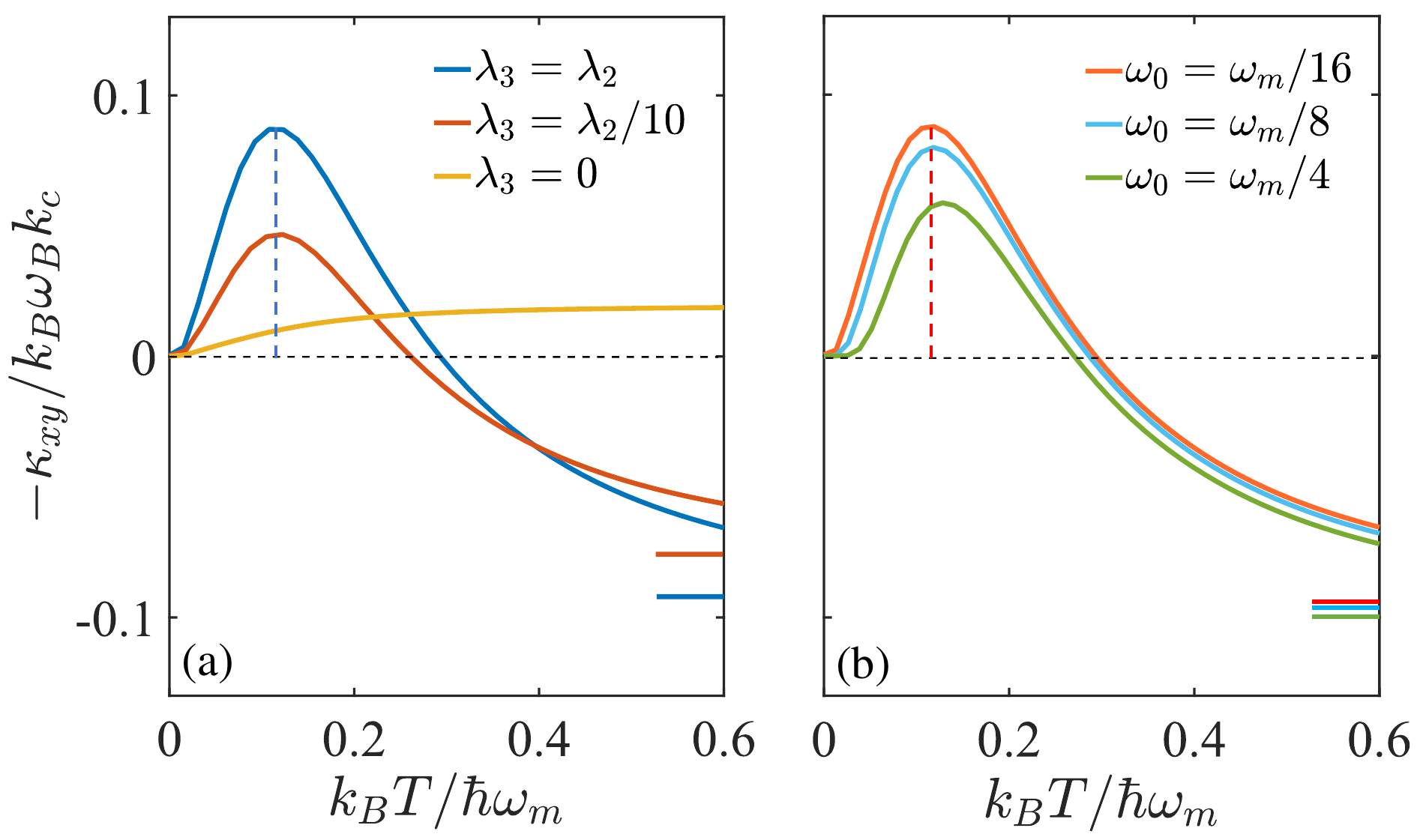}
  \caption{Temperature dependence of the intrinsic thermal Hall conductivity of optical phonons: (a) for different values of $\lambda_{3}$ with $\omega_{m}/\omega_{0}=15$; (b) for various values of $\omega_{m}/\omega_{0}$ with $\lambda_{3}=\lambda_{2}$.  Parameters are $\lambda_{2}=1.5\times10^{4} \text{meV}^{2}\AA^{2}/\hbar^{2}$, $\omega_{B}=5\times 10^{-5}\mathrm{meV}/\hbar$, and $k_{c}=0.8\AA^{-1}$. Horizontal line segments indicate asymptotic values at $T\rightarrow \infty$. Colored dashed lines indicate the positions of the peak temperatures.}
  \label{fig:kappa_t}
\end{figure}

We calculate the intrinsic thermal Hall conductivity using the phonon Hall conductivity formula developed by Qin et al.~\cite{Taoqin2012}:
\begin{align}
\kappa_{xy} & =\hbar\beta k_{B}\int_{0}^{\infty}\mathrm{d}\omega\,\sigma(\omega)\omega^{2}f^{\prime}(\beta\hbar\omega),\label{eq:kxy}\\
\sigma(\omega) & \equiv\sum_{i=\pm}\int\frac{\mathrm{d}^{3}k}{(2\pi)^{3}}\,\Theta\left(\omega-\omega_{\bm{k},i}\right)\Omega_{\bm{k},i},\label{eq:sigma}
\end{align}
where $f(x)$ is the Bose-Einstein distribution function, $\beta\equiv1/k_{B}T$ with $T$ being the temperature, and $\Theta(x)$ denotes the Heaviside function.  Using the Berry curvature determined numerically in Sec.~\ref{sec:berry-curvature-}, we can calculate the thermal Hall conductivity contributed by optical phonons.  For the calculation, we impose the cut-off $|k_{x}|,\,|k_{y}|,\,|k_{z}| \leq k_{c}\sim \pi/a$, where $a$ is the lattice constant of the system.  The cut-off restricts the number of phonon normal modes contributing to the thermal Hall conductivity, which scales as $1/a\sim k_{c}/\pi$ in a three-dimensional system.  The numerical results of the thermal Hall conductivity, in units of $k_{B}\omega_{B}k_{c}$, are presented in Fig.~\ref{fig:kappa_t}. 

From Fig.~\ref{fig:kappa_t}(a), we see that the thermal Hall conductivities of an isotropic system ($\lambda_{3}=0$) and a system with discrete rotational symmetry ($\lambda_{3}\ne 0$) exhibit distinct temperature dependencies. Furthermore, the discrete rotational symmetry significantly enhances the thermal Hall conductivity.  Even a small $\lambda_{3}$ value can induce both the enhancement and the change in the temperature dependence.  This occurs because a small $\lambda_{3}$ value is sufficient to induce the redistribution of the phonon Berry curvature, as pointed out in Sec.~\ref{sec:berry-curvature-}.

The thermal Hall conductivity for $\lambda_{3}\ne 0$ exhibits a non-monotonic temperature dependence, reversing sign and approaching a non-vanishing value at high temperatures.  It peaks at $T_{m}\sim 0.1\hbar\omega_{m}/k_{B}$, where $\omega_{m}=\omega_{\mathrm{d}}(\bm k_{c})$, $\bm k_{c}\equiv (k_{c}, k_{c}, k_{c})$, is the maximum phonon frequency of the transverse modes in the diagonal direction. The scaling relation between the peak temperature and $\omega_{m}$ is insensitive to variations in $\lambda_{3}$, as evident in Fig.~\ref{fig:kappa_t}(a).  The behavior results from competing contributions of the Berry curvatures near the $k_{z}$-axis and the diagonal directions.  The former contributes a negative thermal Hall conductivity and dominates at low temperatures, whereas the latter contributes positively and dominates at high-temperatures. 

The temperature dependence is largely unaffected by variations in $\omega_{0}$, as shown in Fig.~\ref{fig:kappa_t}(b). Both the peak temperature and the high-temperature asymptotic value, when scaled by $\omega_{m}$, exhibit minimal dependence on $\omega_{0}$.  On the other hand, the thermal Hall conductivity is enhanced at low temperatures for smaller $\omega_{0}$, as more phonons can be excited.

Notably, $\kappa_{xy}$ exhibits a non-linear dependence on the magnetic field $B$.  This is evident in Fig.~\ref{fig:kappa_B}, which shows the dependence of $\kappa_{xy}/B$ on the magnetic field.  We see that $\kappa_{xy}/B$ varies linearly with $\ln B$ when $\lambda_{3}\ne 0$, indicating the nonlinear dependence.  In contrast, when $\lambda_{3}=0$, $\kappa_{xy}/B$ remains constant.  The non-linear magnetic field dependence is a result of the singular distribution of the Berry curvature in a system with discrete rotational symmetry, as we will show in the next subsection.

\subsection{Approximate theory} \label{sec:pert-calc}

We can develop an approximate theory to describe the peculiar behavior of the thermal Hall conductivity observed in the numerical results, based on the Berry curvature distribution analyzed in Sec.~\ref{sec:berry-curvature-}.  For simplicity, we consider the contribution from a constant $k_{z}$-plane with $|k_{z}|\gg q_{\mathrm{d}},\,q_{\mathrm{o}}$.  On such a plane, the peaks at $(\pm k_{z}, \pm k_{z}, k_{z})$ and $(0,0,k_{z})$ are well separated  and do not overlap, as illustrated in Fig.~\ref{fig:bc}. The primary contributions to the thermal Hall conductivity from these peaks can be determined using Eqs.~(\ref{eq:kxy}, \ref{eq:sigma}), the approximate expressions for these Berry curvature peaks. 
\begin{figure}[tb]
  \centering
  \includegraphics[width=\columnwidth]{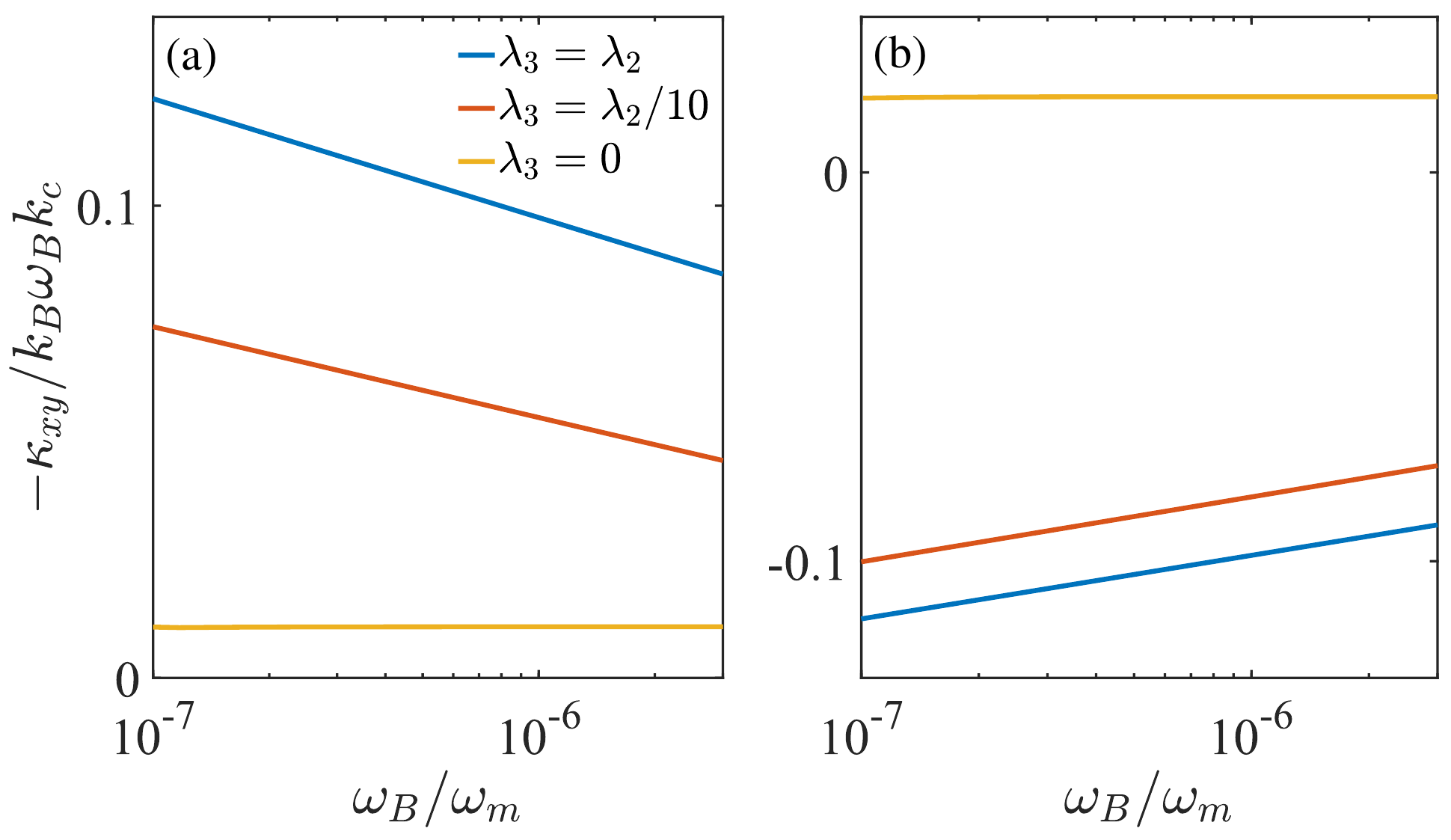}
  \caption{Dependence of the thermal Hall conductivity on the magnetic field $B\propto\omega_{B}$ for various $\lambda_{3}$ values.  The left panel is for temperature $k_{B}T =0.1\hbar\omega_{m}$, near the peaks of $\kappa_{xy}$, and the right panel shows the high-temperature limit.  Other parameters are the same as in Fig.~\ref{fig:kappa_t}.}
  \label{fig:kappa_B}
\end{figure}

We first consider the peaks near $\bm k_{0}\equiv(\pm k_{z},\pm k_{z},k_{z})$.  The partial spectral function $\sigma_{\mathrm{d}}(\omega;k_{z})$, defined similarly to Eq.~\eqref{eq:sigma} but integrated only over the region near $\bm k_{0}$ on the constant $k_{z}$-plane, can be determined using Eqs.~(\ref{eq:dispd}, \ref{eq:bc_R}).  We have (see Appendix~\ref{app:sigma_d}):
\begin{equation}
\sigma_{\mathrm{d}}(\omega;k_{z})=-\frac{\sqrt{3}\omega_{B}}{6\pi\sqrt{[\omega-\omega_{\mathrm{d}}(\bm{k}_{0})]^{2}+4c_{\mathrm{d}}^{2}\omega_{B}^{2}}},\label{eq:r_sigma}
\end{equation}
with $c_{\mathrm{d}}\equiv(3\lambda_{2}+\lambda_{3})/2\sqrt{3}\lambda_{3}$.  Since the approximate expressions for the dispersion and the Berry curvature are valid only in the vicinity of $\bm{k}_{0}$, the result is accurate only within a frequency range, denoted as $[\omega_{l},\omega_{u}]$ with $\omega_{l} < \omega_{\mathrm{d}}(\bm{k}_{0})<\omega_{u}$.

The spectral function produces a thermal Hall conductivity with a nonlinear dependence on    $\omega_{B}$.  To see this, we substitute Eq.~\eqref{eq:r_sigma} into Eq.~\eqref{eq:kxy} and integrate over $[\omega_{l},\omega_{u}]$. After performing integration by parts, we obtain:
\begin{multline}
\kappa_{xy}^{(\mathrm{d})}(k_{z})\approx-\frac{k_{B}\omega_{B}}{\sqrt{3}\pi}\Biggl\{ F\left[\beta\hbar\omega_{\mathrm{d}}(\bm{k}_{0})\right]\ln |\omega_{B}|\\
-\frac{1}{2}\sum_{a=u,l}\Biggl[F(\beta\hbar\omega_{a})\ln\frac{\left|\omega_{\mathrm{d}}(\bm{k}_{0})-\omega_{a}\right|}{c_{\mathrm{d}}}\\
-\int_{\omega_{\mathrm{d}}(\bm{k}_{0})}^{\omega_{a}}\ln\frac{\left|\omega-\omega_{\mathrm{d}}(\bm{k}_{0})\right|}{c_{\mathrm{d}}}\mathrm{d}F(\beta\hbar\omega)\Biggr]\Biggr\},\label{eq:kappa_d}
\end{multline}
with $F(x)\equiv x^{2}/4\sinh^{2}(x/2)$.  To simplify the expression, we exploit the fact that $\omega_{B}$ is much smaller than other frequencies in the expression. Thus, $\omega_{B}$ can be set to zero wherever the substitution does not introduce singularities.  Notably, a singular nonlinear dependence proportional to $\omega_{B}\ln|\omega_{B}|$ arises.
  
The thermal Hall  conductivity from the region near $\bm k_{0}^{\prime}\equiv(0,0,k_{z})$ can be analyzed similarly. The corresponding partial spectral function can be written as (see Appendix~\ref{app:sigma_z}):
\begin{multline}
\sigma_{\mathrm{o}}(\omega;k_{z})=\frac{\omega_{B}\Delta\omega}{4\pi\sqrt{\left[(\Delta\omega)^{2}+c_{\mathrm{o}}^{2}\omega_{B}^{2}\right]\left[(\Delta\omega)^{2}+4c_{\mathrm{o}}^{2}\omega_{B}^{2}\right]}}\\
\times\left[\Theta\left(\Delta\omega+\frac{\omega_{B}}{2}\right)+\Theta\left(\Delta\omega-\frac{\omega_{B}}{2}\right)\right]+\cdots\label{eq:sigma_m}
\end{multline}
where $\Delta\omega\equiv\omega - \omega_{\mathrm{o}}(\bm k_{0}^{\prime})$,  $c_{\mathrm{o}}\equiv(\lambda_{2}+\lambda_{3})/2\lambda_{3}$, and the ellipsis represents terms that are present only when $|\Delta\omega|<\omega_{B}/2$.  This also yields a singular contribution to the thermal Hall conductivity:
\begin{equation}
\kappa_{xy}^{(\mathrm{o})}(k_{z})=\frac{k_{B}\omega_{B}\ln|\omega_{B}|}{2\pi}F\left[\beta\hbar\omega_{\mathrm{o}}(\bm{k}_{0}')\right]+\cdots,\label{eq:kappa_z_kz}
\end{equation}
where the ellipsis denotes terms linear in $\omega_{B}$.

The total thermal Hall conductivity is obtained by summing these contributions after integrating them over $k_{z}\in [-k_{c}, k_{c}]$.  In addition, we also need to include contributions from regions away from the high-symmetry directions, which are linear in $\omega_{B}$.  The final result can be written as:   
\begin{align}
\kappa_{xy} & \approx\mathcal{A}(T)\omega_{B}\ln\left|\frac{\omega_{B}}{\omega_{m}}\right|+\kappa_{xy}^{\prime}(T), \label{eq:kappa_3d}\\
\mathcal{A}(T) & =\frac{k_{B}k_{c}}{\pi^{2}}\left[\frac{\mathcal{F}(\beta\hbar\omega_{m}^{(\mathrm{o})})}{2}-\frac{\mathcal{F}(\beta\hbar\omega_{m})}{\sqrt{3}}\right],\label{eq:Acoef}
\end{align}
where
\begin{equation}
\mathcal{F}(x_{m})\equiv\frac{1}{x_{m}}\int_{0}^{x_{m}}\mathrm{d}y\,\frac{y^{2}}{4\sinh^{2}(y/2)},
\end{equation}
and $\omega_{m}=\omega_{\mathrm{d}}(\bm{k}_{c})$, $\omega_{m}^{(\mathrm{o})}=\omega_{\mathrm{o}}(\bm{k}_{c}')$, with  $\bm{k}_{c}\equiv(k_{c},k_{c},k_{c})$, $\bm{k}_{c}'\equiv(0,0,k_{c})$, are the maximum transverse phonon frequencies along the diagonal and $k_{z}$ directions, respectively.  $\kappa_{xy}^{\prime}(T)$, which cannot be explicitly determined by using only the approximate Berry curvature formulas, accounts for contributions linear in $\omega_{B}$ and high-order nonlinear contributions.  To obtain the formula, we have assumed $\omega_{m},\omega_{m}^{(\mathrm{o})}\gg \omega_{0}$.

\begin{figure}[t]
    \centering
\includegraphics[width=\columnwidth]{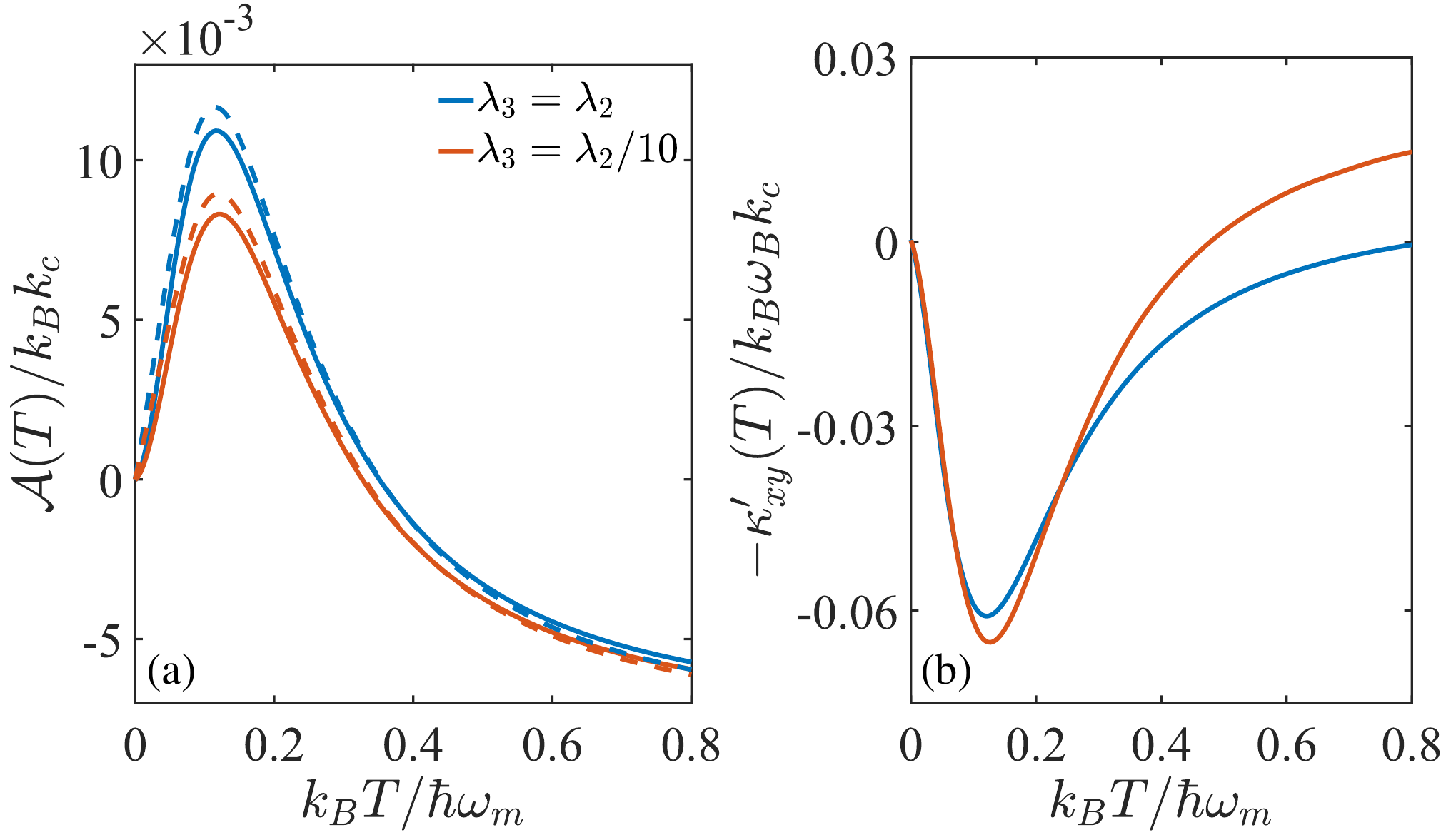}
    \caption{Temperature dependence of (a) $\mathcal{A}(T)$ and (b) $\kappa'_{xy}(T)$. In (a), solid lines represent values inferred from numerical results using linear regressions on the $\kappa_{xy}$ vs. $\omega_{B}$ relations (Fig.~\ref{fig:kappa_B}), while dashed lines show values from the analytic formula Eq.~\eqref{eq:Acoef}.} \label{fig:slope}
\end{figure}

Figure~\ref{fig:slope}(a) shows the temperature dependence of $\mathcal{A}(T)$, which can either be inferred from the numerical results shown in Fig.~\ref{fig:kappa_t} and Fig.~\ref{fig:kappa_B} using the relation Eq.~\eqref{eq:kappa_3d}, or approximately determined using the analytic formula Eq.~\eqref{eq:Acoef}.  The approximate formula aligns closely with the numerical results, indicating that our approximate formula captures the key features of the numerical results.

The analysis elucidates the origin of the non-monotonic temperature dependence of the thermal Hall conductivity.  It arises from competing contributions of different regions near high-symmetry directions in wave-vector space.  At low temperatures, $\kappa_{xy}$ is dominated by the contribution from the region near the $k_{z}$ axis, where optical phonons have lower frequencies, thus $\mathcal{F}(\beta\hbar\omega_{m}^{(\mathrm{o})})\gg\mathcal{F}(\beta\hbar\omega_{m})$.  Conversely, at high temperatures, the contribution from the regions near the diagonal $(\pm1,\pm1,1)$ directions dominates as $\mathcal{F}(\beta\hbar\omega_{m}^{(\mathrm{o})})\approx\mathcal{F}(\beta\hbar\omega_{m})$, and the prefactor of $\mathcal{F}(\beta\hbar\omega_{m})$ ($1/\sqrt{3}$) exceeds that of $\mathcal{F}(\beta\hbar\omega_{m}^{(\mathrm{o})})$ ($1/2$).  The difference in the prefactors can be traced back to the difference in the total Berry phases ($\mp 2\pi$ vs. $\pm\pi \times 4$) as well as the cosine of the angle between the diagonal directions and the $z$-axis ($1/\sqrt{3}$).  Consequently, $\kappa_{xy}$ exhibits a non-monotonic temperature dependence and changes sign with varying temperature.

\section{Summary and discussion}\label{sec:summary}

In summary, we have analyzed the thermal Hall conductivity of optical phonons both numerically and analytically.  The analysis reveals that discrete rotational symmetry, ubiquitous in real materials but overlooked in prior studies, fundamentally alters the band structure and topological features of phonon systems.  It introduces a significant enhancement and a nonlinear $B\ln B$ dependence on the applied magnetic field in the thermal Hall conductivity. 

Although this study focuses on systems with cubic symmetry, our analysis can be generalized straightforwardly to systems with other discrete rotational symmetries.  These systems differ only in the number and orientations of high-symmetry directions.  The primary driving factor, i.e., the concentration of Berry curvatures toward high symmetry directions, remains present.  Thus, their intrinsic thermal Hall conductivities are expected to exhibit qualitatively similar behaviors to those revealed here.

Our analysis, based on a generic effective model for optical phonons, not only offers insights into the qualitative features of the optical phonon contribution to the thermal Hall conductivity but also provides an estimate of its magnitude.  As shown in Fig.~\ref{fig:slope}, the contribution peaks at a temperature $T_{m}\sim 0.1 \hbar\omega_{m}/k_{B}$ with a magnitude of $\sim 0.01 k_{B} k_{c}\omega_{B}\ln|\omega_{B}/\omega_{m}|$, both of which scale with $\omega_{m}$ and are relatively insensitive to the other parameters.  Taking STO as an example, we could map Sr-Ti atoms and O-octahedra to the positive and negative ions in the effective model ~\cite{cowley_lattice_1964}.  We estimate $\omega_{B}\sim5 \times 10^{-5}\,\mathrm{meV}/\hbar$ at $B = 10$ T and $\omega_{m} \approx 20\,\mathrm{meV}/\hbar$~\cite{verdi_quantum_2023}.  Assuming a cutoff $k_{c} \sim \pi/a$ and $a = 3.9 \text{\AA}$~\cite{collignon_metallicity_2019}, the peak thermal Hall conductivity is estimated to be $\sim-10^{-6}\,\mathrm{W/Km}$.  This value is four orders of magnitude smaller than experimental observations ($\sim -80\,\mathrm{mW/Km}$)~\cite{li_phonon_2020}, despite enhancement from the logarithmic factor.  We conclude that the intrinsic mechanism alone cannot explain the experimental data.  The fundamental limit on the magnitude is due to $\omega_{B}$, which is four orders of magnitude smaller than its electron counterpart because the masses of atoms are four orders of magnitude larger than the electron mass.

To resolve the huge discrepancy between the theory and experimental observations, it is necessary to explore more possibilities of enhancing the thermal Hall conductivity.  Within the intrinsic mechanism, our analysis assumes that the coupling between optical phonons and the external magnetic field arises solely from Lorentz forces acting on  charged ions.  However, rigorous theoretical considerations suggest that Berry phases associated with ionic motion may introduce corrections to the coupling~\cite{Taoqin2012,saito_berry_2019,ren_phonon_2021,saparov_lattice_2022}.  Recent numerical investigations indicate that the corrections could enhance the coupling for STO~\cite{zabalo_rotational_2022,haoran},  although the enhancement remains insufficient to explain the observed discrepancy.

On the other hand, a recent experiment seems to indicate that the magnetic moment of optical phonons in STO may be four orders of magnitude larger than previously expected~\cite{basini_terahertz_2024}.  This substantial magnetic moment implies that the coupling strength to the magnetic field, $\omega_{B}$, could be larger by a similar factor.  Although the mechanism behind this phenomenon remains unclear, incorporating it into our model would predict an intrinsic thermal Hall contribution consistent in magnitude with experimental observations.  It also means that the effects predicted in this study would be observable in experiments.

Finally, the fundamental effects of discrete rotational symmetry revealed in this study could also be relevant for extrinsic mechanisms. For example, the skew scattering mechanism~\cite{mori_origin_2014,chen_enhanced_2020,guo_extrinsic_2021, flebus_charged_2022} depends on impurity scattering matrix elements of phonons, which depend on phonon dispersions and states.  It is natural to anticipate that the modifications of phonon bands induced by discrete rotational symmetry will affect phonon scattering.

\begin{acknowledgments}

We thank Jing-Yuan Chen and Haoran Chen for valuable discussions. This work is supported by the National Science Foundation of China under Grant No.~12174005 and the National Key R\&D Program of China under Grand No.~2021YFA1401900.
  
\end{acknowledgments}

\appendix

\section{Effective Hamiltonian}\label{app:H_eff}

We first analysis the region around the point $\bm k_{0}^{\prime}=(0,0,k_{z})$. The eigenvectors of the two degenerate transverse modes at $\bm k_{0}^{\prime}$, $\psi_{i}=\left[\begin{smallmatrix} \bm{u_{i}}\\ -\mathrm{i}\omega_{i}\bm{u}_{i} \end{smallmatrix}\right]$, have the displacement vectors
\begin{equation} \boldsymbol{u}_{1}=\frac{1}{2\omega_{1}}\begin{bmatrix}1\\ -i\\ 0 \end{bmatrix},\,\boldsymbol{u}_{2}=\frac{1}{2\omega_{2}}\begin{bmatrix}1\\ i\\ 0 \end{bmatrix},\label{eq:uo}
\end{equation}
with eigenvalues  $\omega_{1,2}=\omega_{\mathrm{o}}(\bm k_{0}^{\prime})$. The corresponding longitudinal mode has the displacement vector
\begin{equation}
  \boldsymbol{u}_{3}=\frac{1}{\sqrt{2}\omega_{3}}\begin{bmatrix}0\\ 0\\ 1 \end{bmatrix},
\end{equation}
and $\omega_{3}=\omega_{L}(\bm{k}_{0}^{\prime})\equiv\sqrt{\omega_{0}^{2}+\alpha^{2}+(\lambda_{1}+\lambda_{2}+\lambda_{3})k_{z}^{2}}$.  Additionally, there are  three negative-frequency counterparts obtained via $\omega_{i}\rightarrow-\omega_{i}$.  These eigenvectors divide into two subsets: the two positive-frequency transverse modes ($i=1,2$), and the remaining modes, which  include the longitudinal mode ($i=3$) and the three negative-frequency modes.  The effective Hamiltonian Eq.~\eqref{eq:H_eff_G} is constructed  by projecting the full Hamiltonian Eq.~\eqref{Hktilde}  onto the effective subspace spanned by $\{\psi_{1},\psi_{2}\}$, the two positive-frequency transverse modes.

To achieve this, we first express $\tilde{H}_{\bm k}$ for $\bm k = (\delta k_{x},\delta k_{y},k_{z})$ using the eigenvector basis at $\bm k_{0}^{\prime}$.  The $2\times2$ sub-block of the resulting Hamiltonian  matrix for the effective subspace, to second order in $\delta k_{x}$ and $\delta k_{y}$, takes the form:
\begin{equation}
\Tilde{H}_{0}^{(\mathrm{o})}\approx\boldsymbol{\tau}\cdot\boldsymbol{d}^{\prime}(\boldsymbol{k})v_{\mathrm{o}}(k_{z})+\tau_{3}h_{\mathrm{o}}^{\prime}+d^{\prime}_{0}(\bm{k}),\label{eq:H0}
\end{equation}
with
\begin{align}
\boldsymbol{d}^{\prime}(\boldsymbol{k}) & =\frac{1}{\lambda_{3}k_{z}^{2}}\begin{bmatrix}\frac{1}{2}(\delta k_{x}^{2}-\delta k_{y}^{2})[\alpha^{2}+(\lambda_{1}+\lambda_{3})k_{z}^{2}]\\
-\delta k_{x}\delta k_{y}(\alpha^{2}+\lambda_{1}k_{z}^{2})\\
0
\end{bmatrix}^{T},\nonumber \\
d_{0}^{\prime}(\bm{k}) & =\omega_{\mathrm{o}}(\bm{k}_{0}^{\prime})+\frac{[\alpha^{2}+(\lambda_{1}+2\lambda_{2}+\lambda_{3})k_{z}^{2}]}{4\omega_{\mathrm{o}}(\bm{k}_{0}^{\prime})k_{z}^{2}}(\delta k_{x}^{2}+\delta k_{y}^{2}),\nonumber \\
h_{\mathrm{o}}^{\prime} & =-\frac{\omega_{B}}{2}.
\end{align}

The coupling between the effective subspace and the redundant  subspace, which is  spanned by the remaining four modes, introduce a correction to the effective Hamiltonian:
\begin{equation}
\tilde{H}_{1}^{(\mathrm{o})}\approx V\left[\omega_{\mathrm{o}}(\bm{k}_{0}^{\prime})-H_{0}\right]^{-1}V^{\prime},
\end{equation}
where $V$ and $V^{\prime}$ are the sub-blocks of the re-expressed Hamiltonian coupling the effective subspace to the redundant subspace.  To first order in $\delta k_{x}$ and $\delta k_{y}$, the correction is
\begin{align}
V & \approx\frac{\lambda_{1}k_{z}^{2}+\alpha^{2}}{2\sqrt{2}\omega_{L}(\bm{k}_{0}^{\prime})k_{z}}\begin{bmatrix}0 & 0\\
0 & 0\\
\delta k_{x}+i\delta k_{y} & \delta k_{x}-i\delta k_{y}\\
\delta k_{x}+i\delta k_{y} & \delta k_{x}-i\delta k_{y}
\end{bmatrix}^{T},\\
V^{\prime} & =\mathrm{diag}(1,1,1,-1)V^{\dagger},
\end{align}
and $H_{0}$ is the diagonal Hamiltonian matrix at $\bm k_{0}^{\prime}$ for the redundant subspace:
\begin{equation}
H_{0}\approx\text{diag}\left[-\omega_{\mathrm{o}}(\bm{k}_{0}^{\prime}), -\omega_{\mathrm{o}
}(\bm{k}_{0}^{\prime}),  \omega_{L}(\bm{k}_{0}^{\prime}) ,-\omega_{L}(\bm{k}_{0}^{\prime})\right].
\end{equation}
In the limit $\alpha\gg\omega_{\mathrm{o}}(\bm{k})$, we have
\begin{equation}
\tilde{H}_{1}^{\mathrm{(o)}}\approx\boldsymbol{\tau}\cdot\boldsymbol{d}^{\prime\prime}(\boldsymbol{k})v_{\mathrm{o}}(k_{z})+d''_{0}(\bm{k}),\label{eq:Hp}
\end{equation}
with
\begin{eqnarray}
&&\boldsymbol{d}''(\boldsymbol{k})=\frac{\lambda_{3}k_{z}^{2}-\lambda_{1}k_{z}^{2}-\alpha^{2}}{\lambda_{3}k_{z}^{2}}\begin{bmatrix}
    (\delta k_{x}^{2}-\delta k_{y}^{2})/2\\
    -\delta k_{x}\delta k_{y}\\
    0
\end{bmatrix}^{T},\nonumber\\
&&d''_{0}(\bm{k})=\frac{(\lambda_{3}-\lambda_{1})k_{z}^{2}-\alpha^{2}}{4\omega_{\mathrm{o}}(\bm{k}_{0}^{\prime})k_{z}^{2}}(\delta k_{x}^{2}+\delta k_{y}^{2}).
\end{eqnarray}

Summing Eq.~(\ref{eq:H0}) and Eq.~(\ref{eq:Hp}) yields the effective Hamiltonian $\Tilde{H}_{\boldsymbol{k}}^{(\mathrm{o})}$ shown in Eq.~(\ref{eq:H_eff_G}).

The derivation of the effective Hamiltonian Eq.~(\ref{eq:H_eff_R}) for regions near the diagonal direction  follows a similar procedure.  At the point $\bm k_{0}=(k_{z},k_{z},k_{z})$, the eigenvectors of the two degenerate transverse modes have the displacement vectors:
\begin{align}
\boldsymbol{u}_{1} & =\frac{1}{2\sqrt{2}\omega_{1}}\begin{bmatrix}1-\frac{\mathrm{i}}{\sqrt{3}}\\
-1-\frac{\mathrm{i}}{\sqrt{3}}\\
\frac{2\mathrm{i}}{\sqrt{3}}
\end{bmatrix},\,\boldsymbol{u}_{2}=\frac{1}{2\sqrt{2}\omega_{2}}\begin{bmatrix}1+\frac{\mathrm{i}}{\sqrt{3}}\\
-1+\frac{\mathrm{i}}{\sqrt{3}}\\
-\frac{2\mathrm{i}}{\sqrt{3}}
\end{bmatrix},\,\label{eq:ud}
\end{align}
with eigenvalues $\omega_{1,2}=\omega_{\mathrm{d}}(\bm k_{0})$.  To obtain the effective Hamiltonian Eq.~(\ref{eq:H_eff_R}), the full Hamiltonian at $(k_{z}+\delta k_{x},k_{z}+\delta k_{y},k_{z})$ is expanded to linear order in $\delta k_{x}$ and $\delta k_{y}$, and projected onto the subspace spanned by the two eigenvectors.

Unlike Eq.~\eqref{eq:H_eff_G}, redundant modes here do not introduce corrections linear in $\delta k_{x}$ and $\delta k_{y}$.  Their $V$ and $V^{\prime}$ matrices are linear in $\delta k_{x}$ and $\delta k_{y}$, leading to corrections that are at least second order in $\delta k_{x}$ and $\delta k_{y}$.

\section{Partial spectral functions}\label{app:sigma}

\subsection{Diagonal directions}\label{app:sigma_d}

We rewrite the effective Hamiltonian Eq.~(\ref{eq:H_eff_R}) as 
\begin{equation}
\Tilde{H}_{\boldsymbol{k}}^{\mathrm{(d)}}\approx\omega_{\mathrm{d}}(\boldsymbol{k}_{0})+v x+\boldsymbol{\tau}\cdot\boldsymbol{r}+\tau_{3}h_{\mathrm{d}},
\end{equation}
using the coordinate $\boldsymbol{r}=(x, y,0)$ with $x\equiv (\delta k_{x}+\delta k_{y})v_{\mathrm{d}}(k_{z})/\sqrt{6}$, $y\equiv(\delta k_{y}-\delta k_{x})v_{\mathrm{d}}(k_{z})/\sqrt{2}$, and $v\equiv2(3\lambda_{2}+\lambda_{3})/\lambda_{3}$.  The final term in Eq.~(\ref{eq:H_eff_R}) is expanded to first order in $x$ .  The  dispersion relation is given by 
\begin{equation}
\omega_{\pm}(\bm{r})=\omega_{\mathrm{d}}(\bm{k}_{0})+v x\pm\sqrt{x^{2}+ y^{2}+h_{\mathrm{d}}^{2}}.
\end{equation}
The Berry curvature in  $\bm k$-space is related to its counterpart in $\bm r$-space by
\begin{eqnarray}
  \Omega_{\boldsymbol{k},\pm}^{z}=\frac{\partial(x,   y)}{\partial(\delta k_{x},  \delta k_{y})}\Omega_{\boldsymbol{r},\pm}^{z}, 
\end{eqnarray}
with 
\begin{eqnarray}
\Omega_{\boldsymbol{r},\pm}^{z}=\pm\frac{h_{\mathrm{d}}}{2(x^{2}+ y^{2}+h_{\mathrm{d}}^{2})^{3/2}},
\end{eqnarray}
where $\partial (x, y)/\partial(\delta k_{x}, \delta k_{y})$ is the Jacobian determinant of the coordinate transformation.  Consequently, we have the relation
\begin{equation}
\int_{S_{\bm{k}}}\frac{\mathrm{d}^{2}k}{(2\pi)^{2}}\Omega_{\bm{k},\pm}^{z}=-\int_{S_{\bm{r}}}\frac{\mathrm{d}^{2}r}{(2\pi)^{2}}\Omega_{\bm{r},\pm}^{z},
\end{equation}
where $S_{\bm k}$ denotes a region in the $(\delta k_{x}, \delta k_{y})$ plane, mapped to $S_{\bm r}$ in the $xy$-plane, and the negative sign on the right arises due to the Jacobian determinant being negative. 

\begin{figure}[t]
    \centering
\includegraphics[width=0.9\columnwidth]{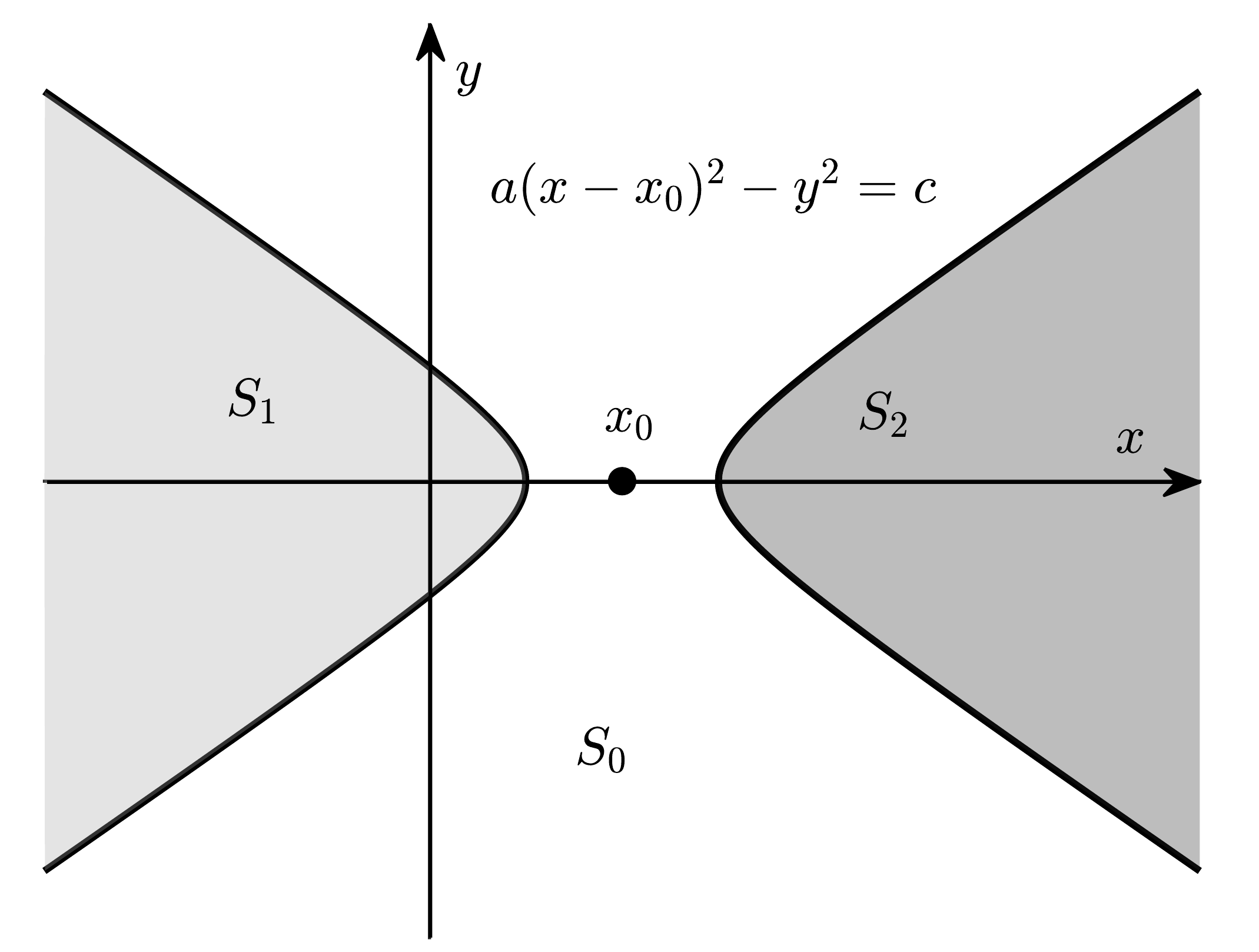}
    \caption{ $xy$-plane is divided into three regions by the hyperbola  $a(x-x_{0})^{2}-y^{2}=c$ with $a\equiv v^{2}-1$, $c\equiv h_{\mathrm{d}}^{2}+\omega^{2}/(v^{2}-1)$, and $x_{0}=v\omega/(v^{2}-1)$.} \label{fig:region}
\end{figure}

The partial spectral function is determined by
\begin{equation}
\sigma_{\mathrm{d}}(\omega;k_{z})=\left[\int_{S_{-}}-\int_{S_{+}}\right]\frac{\mathrm{d}^{2}r}{(2\pi)^{2}}\frac{2h_{\mathrm{d}}}{(x^{2}+ y^{2}+h_{\mathrm{d}}^{2})^{3/2}},
\end{equation}
where $S_{\pm}$ represent regions defined by the conditions $\omega_{\pm}(\bm r) < \omega$.  We have $S_{+}=S_{1}$ and $S_{-}=S_{0}+S_{1}$, where the regions $S_{0}$ and $S_{1}$ are defined in Fig.~\ref{fig:region}.     The integral can be transformed to
\begin{equation}
\sigma_{\mathrm{d}}(\omega;k_{z})=\int_{S_{0}}\frac{\mathrm{d}^{2}r}{(2\pi)^{2}}\frac{2h_{\mathrm{d}}}{(x^{2}+y^{2}+h_{\mathrm{d}}^{2})^{3/2}}.
\end{equation}
It is then straightforward to complete the integral to obtain Eq.~(\ref{eq:r_sigma}).

\subsection{$k_{z}$-direction}\label{app:sigma_z}
The partial spectral function $\sigma_{\mathrm{o}}(\omega;k_{z})$ is the sum of the spectral functions for the upper ($+$) and lower ($-$) bands, defined as
\begin{equation}
\sigma_{\mathrm{o}}^{\pm}(\omega;k_{z})=\int\frac{\mathrm{d}^{2}k}{(2\pi)^{2}}\Theta(\omega-\omega_{\bm{k},\pm})\Omega_{\bm{k},\pm}^{z},\label{eq:sigma_z_part1}
\end{equation}
where $\omega_{\bm k,\pm}$ and $\Omega_{\bm k,\pm}$ are approximated by Eqs.~\eqref{eq:dispz} and \eqref{eq:bc_gamma}, respectively.

For the Berry curvature, the following identity holds:
\begin{equation}
\Omega_{\bm{k},\pm}=\pm\frac{4\omega_{B}}{5+3\cos{4\theta}}\partial_{k^{2}}\left[\frac{1}{\sqrt{d(\boldsymbol{k})^{2}v_{\mathrm{o}}(k_{z})^{2}+h_{\mathrm{o}}^{2}}}\right],
\end{equation}
where $(k,\theta)$ are polar coordinates in the ($k_{x}$, $k_{y}$) plane.

Using this identity, the integral over $k \in [0, k_{\pm}(\theta)]$ can be completed.  Here, $k_{\pm}(\theta)$ is determined by the condition $\omega_{\bm{k},\pm} = \omega$.  The result is  
\begin{multline}
\sigma_{\mathrm{o}}^{\pm}(\omega;k_{z})=\Theta\left(\Delta\omega\mp\frac{\omega_{B}}{2}\right)\int_{0}^{2\pi}\frac{\mathrm{d}\theta}{2\pi^{2}}\,\left[\pm K(\omega,\theta)\right.\\
+\frac{\omega_{B}\Delta\omega}{(\Delta\omega)^{2}(5+3\cos{4\theta})+8c_{\mathrm{o}}^{2}\omega_{B}^{2}}\Biggr],\label{eq:sigma_z_part}
\end{multline}
where $\pm K(\omega, \theta)$ includes terms that alternate in sign for the upper and lower bands.  The contribution from $K(\omega,\theta)$ to $\sigma_{\mathrm{o}}(\omega,k_{z})$ is negligible since its contributions  in the upper and lower bands cancel each other except in the narrow interval with $|\Delta\omega|<\omega_{B}/2$. 

Completing the integral over $\theta$ yields Eq.~(\ref{eq:sigma_m}).

To derive Eq.~\eqref{eq:kappa_z_kz}, we also employ integration by parts and make use of the identity:
\begin{multline}
\int\mathrm{d}\omega\frac{\omega}{\sqrt{\left(\omega^{2}+c_{\mathrm{o}}^{2}\omega_{B}^{2}\right)\left(\omega^{2}+4c_{\mathrm{o}}^{2}\omega_{B}^{2}\right)}}\\
=\ln\left[\sqrt{\omega^{2}+4c_{\mathrm{o}}^{2}\omega_{B}^{2}}+\sqrt{\omega^{2}+c_{\mathrm{o}}^{2}\omega_{B}^{2}}\right].
\end{multline}


\bibliographystyle{apsrev4-2}
\bibliography{main}

\end{document}